\documentclass[preprint,12pt]{elsarticle}
\usepackage[a4paper,left=2.6cm,right=2.6cm,top=2.5cm,bottom=2.5cm]{geometry}
\usepackage{amsmath,amssymb}
\usepackage{graphicx}
\usepackage{booktabs}
\usepackage{array}
\usepackage{tabularx}
\usepackage{longtable}
\usepackage{makecell}
\usepackage{caption}
\usepackage{subcaption}
\usepackage[hidelinks]{hyperref}
\usepackage{xurl}
\usepackage{needspace}
\usepackage{placeins}
\usepackage{float}
\usepackage{microtype}
\DeclareRobustCommand{\rev}[1]{\ifmmode\boldsymbol{#1}\else\textbf{#1}\fi}

\begin{document}

\begin{frontmatter}
\title{A B-factory continuum retune of PYTHIA 8 hadronization parameters using BELLE and BABAR identified-hadron data}

\author[1]{Muhammad Ajaz}
\ead[url]{https://orcid.org/0000-0003-1258-6910}

\author[2]{Haifa I. Alrebdi\corref{cor1}}
\ead{hialrebdi@pnu.edu.sa}
\ead[url]{https://orcid.org/0000-0002-2271-1060}
\cortext[cor1]{Corresponding author}

\address[1]{Department of Physics, Faculty of Science, University of Tabuk, Tabuk 47913, Saudi Arabia}
\address[2]{Department of Physics, College of Science, Princess Nourah bint Abdulrahman University, P.O. Box 84428, Riyadh 11671, Saudi Arabia}

\begin{abstract}
We refine the hadronization sector of a pp PYTHIA~8 tune in $e^+e^-\to q\bar q$ production using selected BELLE and BABAR measurements near the $\Upsilon(4S)$ region. The study is performed with PYTHIA~8.316 and Rivet~4.1.1 and is restricted to parameters used in this $e^+e^-$ setup. Starting from the five-parameter hadronization subset of the pp tune, we carry out a staged extension guided by the remaining differences in the BELLE and BABAR data. The final comparison uses a fixed common set of 20,803 bins and samples of 1,000,000 generated events per analysis and tune point. On this basis, the selected refined tune gives a bin-weighted score of 73.42, compared with 76.49 for the Skands $e^+e^-$ reference and 79.22 for Monash~2013 $e^+e^-$. It remains the best-scoring tune for the BELLE charged-hadron, baryon, and single- and dihadron measurements, while Skands still performs better for the BABAR charged-hadron sample and the BELLE meson sample. The bin-weighted ordering is driven primarily by the BELLE 2020 sample. The refined tune gives a small but stable improvement for the selected BELLE and BABAR measurements, although clear differences between the individual datasets remain.
\end{abstract}


\end{frontmatter}

\section{Introduction}
\label{sec:introduction}
The pp tune reported in Ref.~\cite{Alrebdi:2026ajaz} was constructed as a sequential retune of PYTHIA~8.316 \cite{Sjostrand:2014zea,Bierlich:2022pfr} around the Monash~2013 baseline for soft-QCD observables in proton--proton collisions. A separate pp validation against ATLAS underlying-event and azimuthal-ordering data at $\sqrt{s}$ = 0.9 and 7 TeV showed improvement in several measurement groups \cite{Ajaz:2026UEAzimuth}. A useful test is to determine whether the non-perturbative parameter values obtained from pp collisions also describe $e^+e^-$ data. In $e^+e^-$ annihilation there are no beam remnants and no multiparton interactions, so the comparison is driven directly by fragmentation, flavour production, and baryon formation. The B-factory data near the $\Upsilon(4S)$ therefore provide a direct test of how far a pp-derived non-perturbative tune can be used in $e^+e^-$ collisions before a dedicated channel-specific refinement becomes necessary. Previous pp studies show that the performance of PYTHIA depends on the observable, particle species, and kinematic region~\cite{Waqar:2024cpc,Alrebdi:2025epjpUE,Waqar:2025epjpHadron}. This motivates a separate test of the pp-derived parameters using $e^+e^-$ data.

The available measurements are now broad enough to make that comparison meaningful. Near $\sqrt{s}=10.52$--10.58~GeV, BELLE and BABAR provide identified charged-hadron spectra, baryon production spectra, single- and dihadron distributions, and light- and charm-meson spectra \cite{Leitgab:2013qh,Lees:2013rqd,Niiyama:2017wpp,Seidl:2020mqc,Belle:2024vua}. These measurements constrain distinct parts of the string-fragmentation model~\cite{Andersson:1983ia} while remaining close enough in energy that the comparison is not obscured by a large beam-energy lever arm. They therefore offer a compact but stringent basis for an $e^+e^-$ tune refinement. The starting point of the present $e^+e^-$ study is the nine-parameter pp tune of Ref.~\cite{Alrebdi:2026ajaz}, evaluated in a channel where colour reconnection and multiparton interactions are not used in this setup. In practice, this reduces the working starting point to a five-parameter hadronization subset. That subset gives competitive scores for several selected B-factory observables, but it does not surpass Monash on the combined \(e^+e^-\) scoring basis. The purpose of the present paper is therefore to document a dedicated $e^+e^-$ refinement of that baseline and to compare the resulting tune with the two relevant references, Monash~2013 $e^+e^-$~\cite{Skands:2014pea} and the older internal PYTHIA $e^+e^-$ tune implemented by \nolinkurl{Tune:ee = 4}~\cite{Bierlich:2022pfr}.

\section{Experimental basis and generator configuration}
\label{sec:setup}

\subsection{Experimental data set}

The comparison is based on five Rivet analyses from the BELLE and BABAR B-factory measurements. They cover identified charged hadrons at 10.52 and 10.54~GeV, baryon production at 10.52~GeV, the large BELLE single- and dihadron sample at 10.58~GeV, and the recent BELLE meson spectrum analysis at 10.58~GeV. The selected measurements were not used in the pp tune construction of Ref.~\cite{Alrebdi:2026ajaz}, so the $e^+e^-$ study is independent by construction from the original tuning basis.

The measurements are close enough in energy to be compared together, but they constrain different parts of the model. BELLE~2013 provides the 10.52~GeV charged-hadron reference measurement, BABAR~2013 supplies an independent experimental cross-check at 10.54~GeV, BELLE~2017 constrains the baryon sector, BELLE~2020 dominates the total number of bins and therefore the final bin-weighted score, and BELLE~2025 tests meson production in a sector that proved especially informative during the final refinement. The full analysis inventory is listed in Table~\ref{tab:datasets}.

{\small
\setlength{\LTleft}{0pt}
\setlength{\LTright}{0pt}
\setlength{\tabcolsep}{4pt}
\renewcommand{\arraystretch}{1.08}

\begin{longtable}{@{}
>{\raggedright\arraybackslash}p{0.23\textwidth}
>{\centering\arraybackslash}p{0.10\textwidth}
>{\raggedright\arraybackslash}p{0.16\textwidth}
>{\raggedright\arraybackslash}p{0.17\textwidth}
>{\raggedright\arraybackslash}p{0.26\textwidth}
@{}}
\caption{Continuum analyses used in the $e^+e^-$ refinement and final comparison.}
\label{tab:datasets}\\
\toprule
Analysis & $\sqrt{s}$ [GeV] & Generator-frame beam energies & Observable family & Dataset note \\
\midrule
\endfirsthead

\multicolumn{5}{@{}l}{\textit{Table \thetable\ continued.}}\\
\toprule
Analysis & $\sqrt{s}$ [GeV] & Generator-frame beam energies & Observable family & Dataset note \\
\midrule
\endhead

\midrule
\multicolumn{5}{r@{}}{\textit{Continued on next page}}\\
\endfoot

\bottomrule
\endlastfoot

\nolinkurl{BELLE_2013_I1216515} & 10.52 & asymmetric (3.50 + 7.91) & identified charged hadrons & Independent of the pp tuning basis. Used as a 10.52~GeV cross-check, although \nolinkurl{BELLE_2020_I1777678} supersedes part of the single-hadron program. \\

\nolinkurl{BELLE_2017_I1606201} & 10.52 & symmetric (5.26 + 5.26) & baryon spectra & Independent of the pp tuning basis. \\

\nolinkurl{BABAR_2013_I1238276} & 10.54 & asymmetric (3.50 + 8.00) & identified charged hadrons & Independent measurement at a nearby energy; included as a cross-check on the charged-hadron sector. \\

\nolinkurl{BELLE_2020_I1777678} & 10.58 & symmetric (5.29 + 5.29) & single- and dihadron spectra & Highest-statistics and highest-granularity sample in the suite; it supplies most bins in the common comparison. \\

\nolinkurl{BELLE_2025_I2849895} & 10.58 & symmetric (5.29 + 5.29) & meson spectra & Recent Rivet routine for light- and charm-meson production. \\

\end{longtable}
}

\subsection{Generator setup and comparison points}

All events were generated with PYTHIA~8.316 and analysed with Rivet~4.1.1~\cite{Bierlich:2020skw}. The common $e^+e^-$ production card uses the direct Rivet interface with \nolinkurl{WeakSingleBoson:ffbar2gmZ = on}, \nolinkurl{23:onMode = off}, and \nolinkurl{23:onIfAny = 1 2 3 4 5}. Lepton PDFs are disabled, multiparton interactions are switched off, and hadronization is left fully active. Where symmetric beam energies are listed, event generation was performed in the centre-of-mass frame. The compared Rivet observables are defined in that frame and do not require applying the experimental laboratory boost. Independent random seeds were used for all statistically separate event samples. The exact run conditions are summarized in Table~\ref{tab:run-conditions}.

\begin{table}[H]
\centering
\small
\caption{Generator settings for the final $e^+e^-$ comparison.}
\label{tab:run-conditions}

\setlength{\tabcolsep}{4pt}
\renewcommand{\arraystretch}{1.08}

\begin{tabular}{@{}
>{\raggedright\arraybackslash}p{0.20\textwidth}
>{\raggedright\arraybackslash}p{0.66\textwidth}
@{}}
\toprule
Item & Setting \\
\midrule
Generator
& PYTHIA~8.316  \\

Analysis framework
& Rivet~4.1.1 \\

Hard process
& \nolinkurl{WeakSingleBoson:ffbar2gmZ = on}, \nolinkurl{23:onMode = off}, and \nolinkurl{23:onIfAny = 1 2 3 4 5} \\

Beam species
& $e^{+}e^{-}$, with point-specific beam energies listed in Table~\ref{tab:datasets} \\

Underlying event
& \nolinkurl{PartonLevel:MPI = off} \\

Hadronization
& \nolinkurl{HadronLevel:all = on} \\

Statistics
& Statistically independent samples with 1,000,000 generated events per analysis and tune point in the final comparison \\

Tune points
& Refined tune, Monash~2013 $e^{+}e^{-}$, and Skands $e^{+}e^{-}$ (\nolinkurl{Tune:ee = 4}) \\

Analyses
& \nolinkurl{BELLE_2013_I1216515}, \nolinkurl{BELLE_2017_I1606201}, \nolinkurl{BABAR_2013_I1238276}, \nolinkurl{BELLE_2020_I1777678}, and \nolinkurl{BELLE_2025_I2849895} \\

Common bin set
& 20,803 bins common to all final tune comparisons \\

\bottomrule
\end{tabular}
\end{table}

The final comparison uses three tune points: the refined tune, Monash~2013
\(e^+e^-\), and the Skands \(e^+e^-\) reference. Their parameter
values are listed in Table~\ref{tab:tune-parameters}. The refined tune is
defined relative to the Monash \(e^+e^-\) baseline, \nolinkurl{Tune:ee = 7},
and only varies parameters that are active in the leptonic setup. The inactive
pp-specific colour-reconnection and multiparton-interaction parameters are
therefore not retuned. This keeps the comparison restricted to the
hadronization sector relevant for the $e^+e^-\to q\bar q$ final state.

{\small
\setlength{\LTleft}{0pt}
\setlength{\LTright}{0pt}
\setlength{\tabcolsep}{4pt}
\renewcommand{\arraystretch}{1.08}

\begin{longtable}{@{}
>{\raggedright\arraybackslash}p{0.36\textwidth}
>{\raggedleft\arraybackslash}p{0.18\textwidth}
>{\raggedleft\arraybackslash}p{0.18\textwidth}
>{\raggedleft\arraybackslash}p{0.16\textwidth}
@{}}
\caption{Final tune parameters used in the comparison of $e^+e^-$ collisions at the BELLE/BABAR energies. Entries marked ``default'' denote values not explicitly retuned in the corresponding reference configuration. The refined-tune values are reported to the numerical precision supported by the one-parameter profile study. The exact values used in event generation are retained in the input tune card.}
\label{tab:tune-parameters}\\
\toprule
Parameter & Refined tune & Monash~2013 & Skands $e^{+}e^{-}$ \\
\midrule
\endfirsthead

\multicolumn{4}{@{}l}{\textit{Table \thetable\ continued.}}\\
\toprule
Parameter & Refined tune & Monash~2013 & Skands $e^{+}e^{-}$ \\
\midrule
\endhead

\midrule
\multicolumn{4}{r@{}}{\textit{Continued on next page}}\\
\endfoot

\bottomrule
\endlastfoot

\nolinkurl{Tune:ee}                  & 7      & 7      & 4 \\
\nolinkurl{StringZ:aLund}            & 0.75 & 0.680  & default \\
\nolinkurl{StringZ:bLund}            & 1.0 & 0.980  & default \\
\nolinkurl{StringFlav:probStoUD}     & 0.21 & 0.217  & default \\
\nolinkurl{StringPT:sigma}           & 0.32 & 0.335  & default \\
\nolinkurl{StringFlav:probQQtoQ}     & 0.092 & 0.081  & default \\
\nolinkurl{StringFlav:probSQtoQQ}    & 0.77 & default & default \\
\nolinkurl{StringFlav:probQQ1toQQ0}  & 0.042 & default & default \\
\nolinkurl{StringFlav:mesonUDvector} & 0.58 & default & default \\
\nolinkurl{StringFlav:mesonSvector}  & 0.54 & default & default \\
\nolinkurl{StringFlav:mesonCvector}  & 0.82 & default & default \\
\nolinkurl{StringFlav:etaSup}        & 0.68 & default & default \\
\nolinkurl{StringFlav:etaPrimeSup}   & 0.26 & default & default \\
\nolinkurl{StringFlav:popcornRate}   & 0.32 & default & default \\

\end{longtable}
}

\section{Sequential retuning strategy}
\label{sec:strategy}

The retuning was carried out in a staged sequence rather than in a single high-dimensional scan. This follows the usual generator-tuning practice of using controlled parameter variations and validation samples rather than interpreting a single scan point in isolation~\cite{Buckley:2009bj}. This staged procedure also allows the effects of the added parameter groups to be examined separately. Table~\ref{tab:retuning-sequence} summarizes the sequence, and Table~\ref{tab:retune-progress} gives the stagewise equal-analysis performance for the fixed-mask validation comparison based on 200,000 events per analysis point.

The starting point was the nine-parameter pp tune of Ref.~\cite{Alrebdi:2026ajaz}, evaluated in the $e^+e^-$ channel. Since \nolinkurl{ColourReconnection:range}, \nolinkurl{MultipartonInteractions:pT0Ref}, \nolinkurl{MultipartonInteractions:ecmPow}, and \nolinkurl{MultipartonInteractions:expPow} are inactive in this setup, the effective baseline reduces to a five-parameter hadronization subset defined by \nolinkurl{StringZ:aLund}, \nolinkurl{StringZ:bLund}, \nolinkurl{StringFlav:probStoUD}, \nolinkurl{StringPT:sigma}, and \nolinkurl{StringFlav:probQQtoQ}. This baseline remained competitive across several B-factory observables, but it still trailed Monash slightly in the combined score.

The first dedicated extension therefore opened four flavour-sensitive parameters with a direct hadronization interpretation: \nolinkurl{StringFlav:probSQtoQQ}, \nolinkurl{StringFlav:probQQ1toQQ0}, \nolinkurl{StringFlav:mesonUDvector}, and \nolinkurl{StringFlav:mesonSvector}. This step improved the combined $e^+e^-$ score relative to both references, but the remaining differences were concentrated in BABAR~2013 and BELLE~2025.

The final refinement was constructed around the remaining BABAR~2013 and BELLE~2025 deficits. A 25-point scan based on Latin-hypercube sampling~\cite{McKay:1979lhd} was centered on the leading point from the previous step, with the nine already active parameters varied in a narrow range and with \nolinkurl{StringFlav:mesonCvector}, \nolinkurl{StringFlav:etaSup}, \nolinkurl{StringFlav:etaPrimeSup}, and \nolinkurl{StringFlav:popcornRate}, which controls popcorn baryon production in the string picture~\cite{Eden:1996bk}, added to the scan. In the high-statistics stage, 48 parameter vectors were examined with 200,000 generated events per analysis. The best points were then compared with the reference tunes using 1,000,000 generated events per analysis.

For each bin, the contribution to the comparison score is
\[
\chi_i^2=\frac{\left(y_i^{\mathrm{MC}}-y_i^{\mathrm{data}}\right)^2}{\sigma_{i,\mathrm{data}}^2+\sigma_{i,\mathrm{MC}}^2}.
\]
Here \(i\) labels a bin in the fixed common mask \(\mathcal{M}\), \(y_i^{\mathrm{MC}}\) and \(y_i^{\mathrm{data}}\) are the simulated and reference bin contents, and \(\sigma_{i,\mathrm{MC}}\) and \(\sigma_{i,\mathrm{data}}\) are the corresponding diagonal uncertainties. The three comparison scores are
\[
S_{\mathrm{bin}}=\frac{1}{N_{\mathrm{bin}}}\sum_{i\in\mathcal{M}}\chi_i^2,\qquad N_{\mathrm{bin}}=|\mathcal{M}|=20,803,
\]
\[
S_{\mathrm{analysis}}=\frac{1}{5}\sum_{a\in\mathcal{A}}S_a,\qquad S_a=\frac{1}{N_a}\sum_{i\in\mathcal{M}_a}\chi_i^2,
\]
\[
S_{\mathrm{family}}=\frac{1}{8}\sum_{f\in\mathcal{F}}S_f,\qquad S_f=\frac{1}{N_f}\sum_{i\in\mathcal{M}_f}\chi_i^2.
\]
The set \(\mathcal{A}\) contains the five analyses listed in Table~\ref{tab:datasets}. The eight equally weighted observable families are the BELLE~2013 charged-hadron spectra, the BABAR~2013 charged-hadron spectra, the BELLE~2017 strange-baryon spectra, the BELLE~2017 charmed-baryon spectra, the BELLE~2020 single-hadron spectra, the BELLE~2020 dihadron spectra, the BELLE~2025 light-meson spectra, and the BELLE~2025 charm-meson spectra. The score uses the uncertainty provided with each Rivet/YODA reference bin and the Monte Carlo statistical uncertainty. The input does not separate experimental statistical and systematic components. For asymmetric (Y) errors, the larger absolute up/down error is used. No normalization correction is fitted, and full covariance matrices are not available. Monte Carlo statistical uncertainties are included in the denominator, following the standard need to account for finite simulated samples~\cite{Barlow:1993dm}. These scores are therefore used to compare tune points and are not interpreted as formal goodness-of-fit probabilities.

\begin{table}[H]
\centering
\small
\caption{Sequential construction of the selected $e^+e^-$ tune.}
\label{tab:retuning-sequence}

\setlength{\tabcolsep}{2pt}
\renewcommand{\arraystretch}{1.08}

\begin{tabularx}{\textwidth}{@{}
>{\raggedright\arraybackslash}p{0.13\textwidth}
>{\raggedright\arraybackslash}p{0.34\textwidth}
>{\raggedright\arraybackslash}p{0.23\textwidth}
X@{}}
\toprule
Stage & Parameters varied & Statistics & Result \\
\midrule

pp-derived active subset
&
Five hadronization parameters from Ref.~\cite{Alrebdi:2026ajaz}: 
\nolinkurl{aLund}, \nolinkurl{bLund}, \nolinkurl{probStoUD}, 
\nolinkurl{sigma}, and \nolinkurl{probQQtoQ}. 
The pp-specific CR/MPI parameters are not used in the $e^+e^-$ setup.
&
Initial $e^{+}e^{-}$ benchmark comparison.
&
Competitive for BELLE~2013 and BELLE~2017, but below Monash in the combined score.
\\

First $e^{+}e^{-}$ extension
&
Added \nolinkurl{probSQtoQQ}, \nolinkurl{probQQ1toQQ0}, 
\nolinkurl{mesonUDvector}, and \nolinkurl{mesonSvector}.
&
Scouting runs followed by 200,000 events per analysis for the leading points.
&
Improved the combined score, while BABAR~2013 and BELLE~2025 remained less well described.
\\

Final refinement step
&
Added \nolinkurl{mesonCvector}, \nolinkurl{etaSup}, 
\nolinkurl{etaPrimeSup}, and \nolinkurl{popcornRate}; 25 Latin-hypercube points were centred on the leading region.
&
48 parameter vectors with 200,000 events per analysis; final comparison with 1,000,000 events per analysis.
&
The selected tune gives the lowest bin-weighted, equal-analysis, and family-balanced scores among the final compared tunes.
\\

\bottomrule
\end{tabularx}
\end{table}

\begin{table}[H]
\centering
\small
\caption{Stagewise equal-analysis performance in the fixed-common-bin validation comparison with 200,000 generated events per analysis point.}
\label{tab:retune-progress}

\setlength{\tabcolsep}{2pt}
\renewcommand{\arraystretch}{1.10}

\begin{tabular}{@{}
>{\raggedright\arraybackslash}p{0.18\textwidth}
>{\raggedright\arraybackslash}p{0.40\textwidth}
>{\centering\arraybackslash}p{0.16\textwidth}
>{\centering\arraybackslash}p{0.18\textwidth}
@{}}
\toprule
Configuration
&
Additional active parameters beyond the pp-derived active subset
&
\shortstack{Equal-analysis\\score}
&
\shortstack{\(\Delta S_{\mathrm{analysis}}\)\\from final refinement}
\\
\midrule

pp-derived active subset
&
none
&
76.246
&
7.534
\\

First \(e^{+}e^{-}\) extension
&
\nolinkurl{probSQtoQQ},
\nolinkurl{probQQ1toQQ0},
\nolinkurl{mesonUDvector},
\nolinkurl{mesonSvector}
&
74.968
&
6.256
\\

Final refinement step
&
\nolinkurl{mesonCvector},
\nolinkurl{etaSup},
\nolinkurl{etaPrimeSup},
\nolinkurl{popcornRate}
&
68.712
&
0.000
\\

Skands \(e^{+}e^{-}\)
&
PYTHIA \(e^{+}e^{-}\) reference tune
&
71.334
&
2.623
\\

Monash~2013 \(e^{+}e^{-}\)
&
Monash~2013 reference tune
&
74.427
&
5.716
\\

\bottomrule
\end{tabular}
\end{table}

\section{Results}
\label{sec:results}

\subsection{Global and point-level comparison}

The final three-way comparison is summarized in Table~\ref{tab:global-summary} and Table~\ref{tab:point-summary}. In the one-million-event comparison, the selected refined tune gives a bin-weighted score of 73.42, compared with 76.49 for Skands and 79.22 for Monash. The same tune is also lowest for the equal-analysis and observable-family-balanced scores.

The point-level scores show a non-uniform pattern. The refined tune gives the
lowest score for BELLE~2013, BELLE~2017, and BELLE~2020, with
scores of \(245.94\), \(98.09\), and \(73.28\), respectively.
Skands gives the lowest score for BABAR~2013 and BELLE~2025, with
scores of \(24.36\) and \(29.94\), compared with \(30.90\)
and \(40.04\) for the refined tune. The refined tune therefore changes the
balance among the selected data sets rather than improving every observable
family.

\begin{table}[H]
\centering
\caption{Global continuum comparison of the three final $e^+e^-$ tune configurations.}
\label{tab:global-summary}

\begin{tabular}{@{}lrrrrr@{}}
\toprule
Tune & $\chi^2$ & $N_{\mathrm{bin}}$ & Bin-weighted & Equal-analysis & Family-balanced \\
\midrule
Refined tune    & 1527269.7 & 20803 & 73.42 & 97.65 & 81.72 \\
Monash~2013     & 1648042.3 & 20803 & 79.22 & 107.98 & 89.95 \\
Skands $e^{+}e^{-}$  & 1591323.1 & 20803 & 76.49 & 102.95 & 84.26 \\
\bottomrule
\end{tabular}
\end{table}

\begin{table}[H]
\centering
\small
\caption{Point-by-point scores for the five BELLE/BABAR analyses used in the final three-tune comparison.}
\label{tab:point-summary}

\setlength{\tabcolsep}{4pt}
\renewcommand{\arraystretch}{1.08}

\begin{tabular}{@{}
>{\raggedright\arraybackslash}p{0.25\textwidth}
>{\raggedright\arraybackslash}p{0.18\textwidth}
r
r
r
r
>{\raggedright\arraybackslash}p{0.10\textwidth}
@{}}
\toprule
Analysis & Family & Refined tune & Monash~2013 & Skands $e^{+}e^{-}$ & $N_{\mathrm{bin}}$ & Best tune \\
\midrule
\nolinkurl{BELLE_2013_I1216515} & identified charged hadrons   & 245.94 & 290.45 & 276.96 &   156 & Refined tune \\
\nolinkurl{BELLE_2017_I1606201} & baryon spectra               &  98.09 &  106.17 &  106.92 &   266 & Refined tune \\
\nolinkurl{BABAR_2013_I1238276} & identified charged hadrons   &  30.90 &  25.25 &  24.36 &   270 & Skands $e^{+}e^{-}$ \\
\nolinkurl{BELLE_2020_I1777678} & single- and dihadron spectra &  73.28 &  79.11 &  76.58 & 19530 & Refined tune \\
\nolinkurl{BELLE_2025_I2849895} & meson spectra                &  40.04 &  38.91 &  29.94 &   581 & Skands $e^{+}e^{-}$ \\
\bottomrule
\end{tabular}
\end{table}

\subsection{Selected comparison panels}

\begin{figure}[p]
\centering
\begin{subfigure}[t]{0.48\textwidth}
\includegraphics[width=\linewidth]{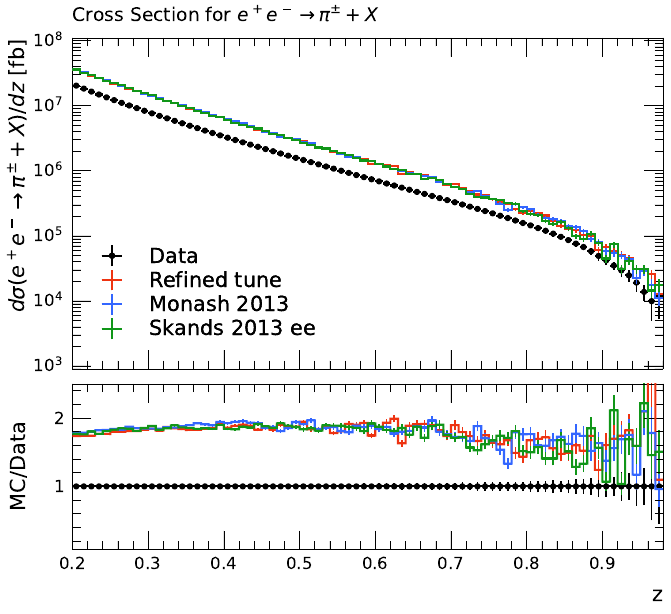}
\caption{BELLE~2013 charged-pion spectrum at 10.52~GeV.}
\end{subfigure}\hfill
\begin{subfigure}[t]{0.48\textwidth}
\includegraphics[width=\linewidth]{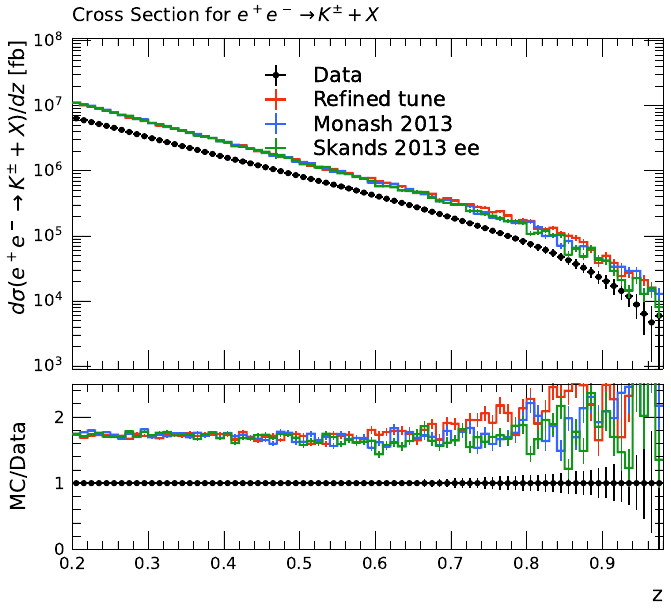}
\caption{BELLE~2013 charged-kaon spectrum at 10.52~GeV.}
\end{subfigure}
\vspace{0.8em}
\begin{subfigure}[t]{0.48\textwidth}
\includegraphics[width=\linewidth]{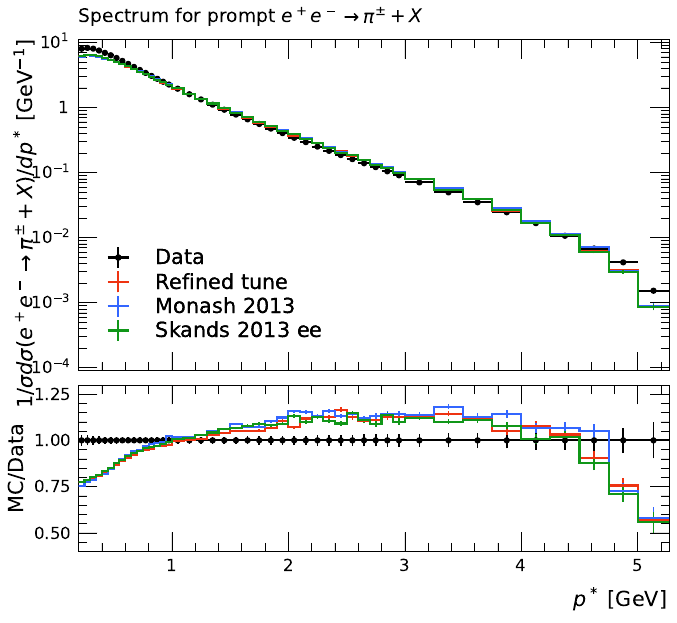}
\caption{BABAR prompt charged-pion spectrum at 10.54~GeV.}
\end{subfigure}\hfill
\begin{subfigure}[t]{0.48\textwidth}
\includegraphics[width=\linewidth]{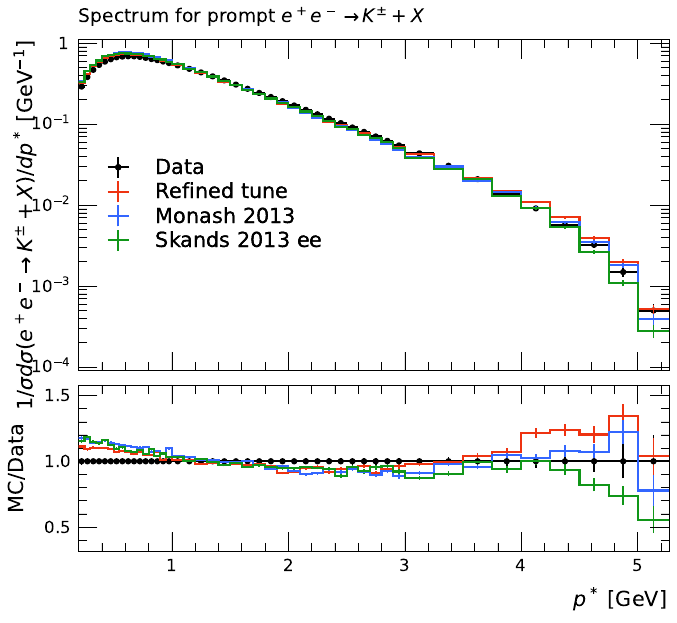}
\caption{BABAR prompt charged-kaon spectrum at 10.54~GeV.}
\end{subfigure}\hfill
\begin{subfigure}[t]{0.48\textwidth}
\includegraphics[width=\linewidth]{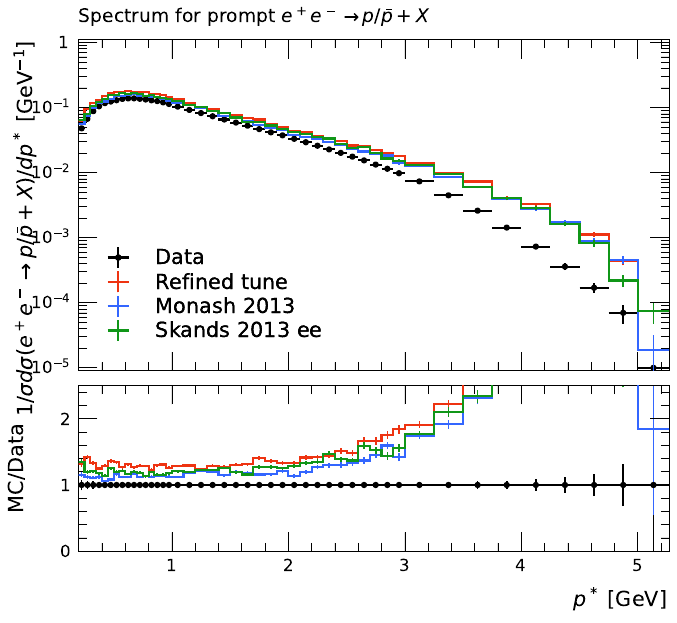}
\caption{BABAR prompt proton spectrum at 10.54~GeV.}
\end{subfigure}
\caption{Low-energy charged-hadron sector. The selected refined tune remains best on the BELLE~2013 charged-hadron spectra, although the nearby BABAR sample is still described more accurately by the Skands reference.}
\label{fig:low-energy-panels}
\end{figure}

\begin{figure}[p]
\centering
\begin{subfigure}[t]{0.48\textwidth}
\includegraphics[width=\linewidth]{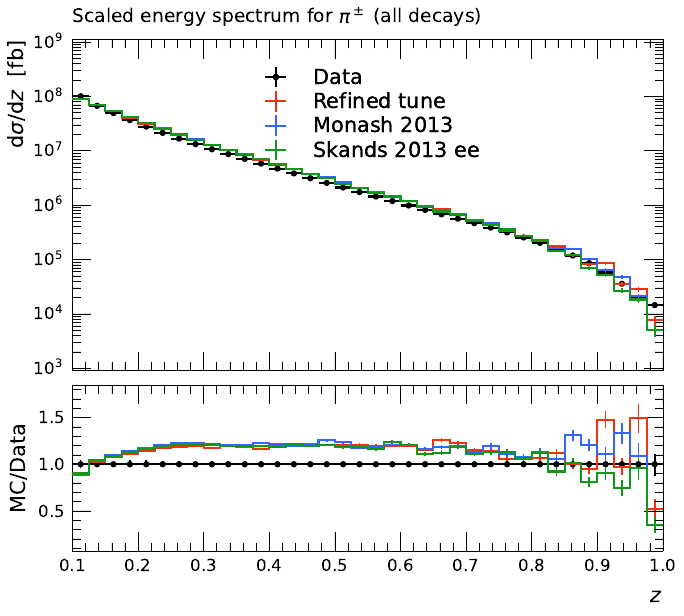}
\caption{Single-hadron $\pi^\pm$ spectrum.}
\end{subfigure}\hfill
\begin{subfigure}[t]{0.48\textwidth}
\includegraphics[width=\linewidth]{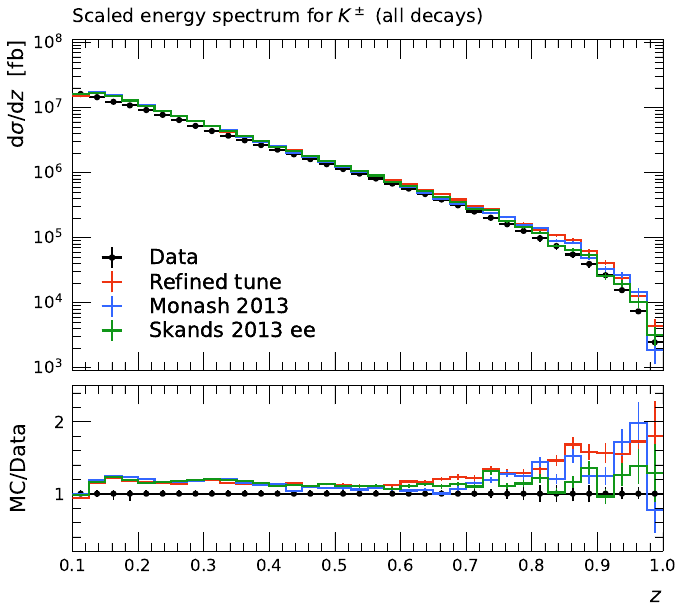}
\caption{Single-hadron $K^\pm$ spectrum.}
\end{subfigure}
\vspace{0.8em}
\begin{subfigure}[t]{0.48\textwidth}
\includegraphics[width=\linewidth]{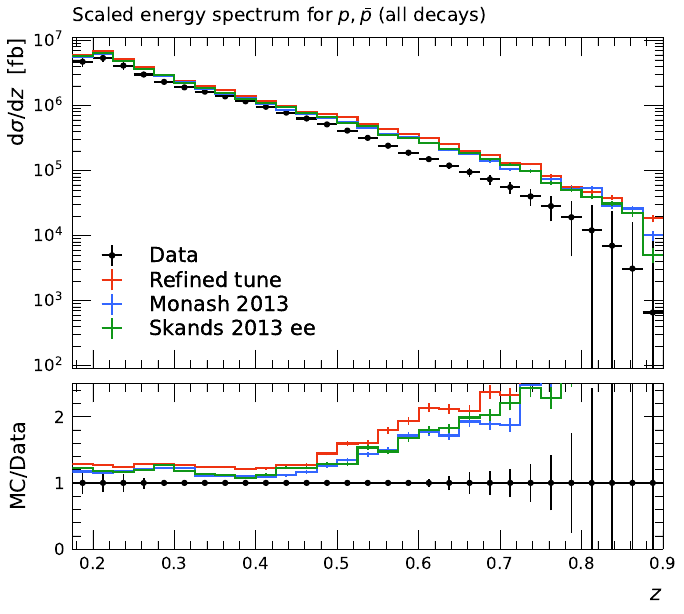}
\caption{Single-hadron $p,\bar p$ spectrum.}
\end{subfigure}\hfill
\begin{subfigure}[t]{0.48\textwidth}
\includegraphics[width=\linewidth]{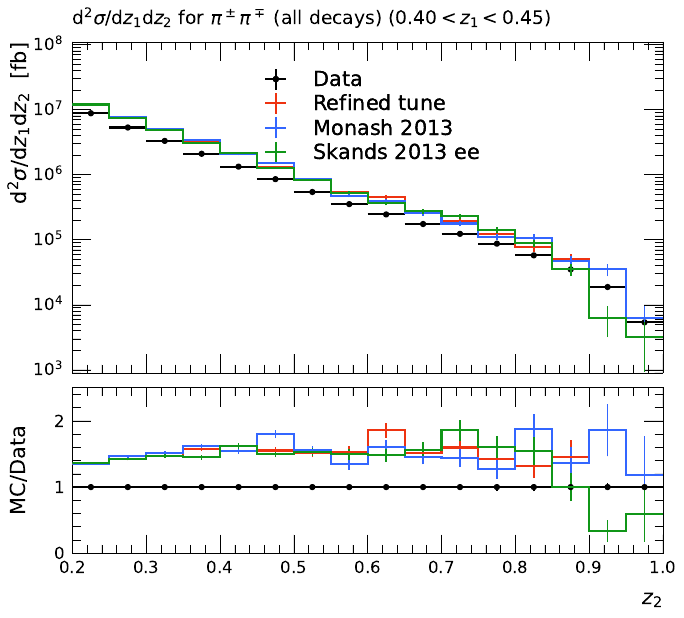}
\caption{Dihadron $\pi^\pm\pi^\mp$ spectrum for $0.40<z_1<0.45$.}
\end{subfigure}
\caption{Selected BELLE~2020 single- and dihadron spectra. The differences between the three tune curves are modest, but BELLE~2020 supplies most of the common bins and therefore largely determines the bin-weighted ranking.}

\label{fig:belle2020-panels}
\end{figure}

\begin{figure}[p]
\centering
\begin{subfigure}[t]{0.48\textwidth}
\includegraphics[width=\linewidth]{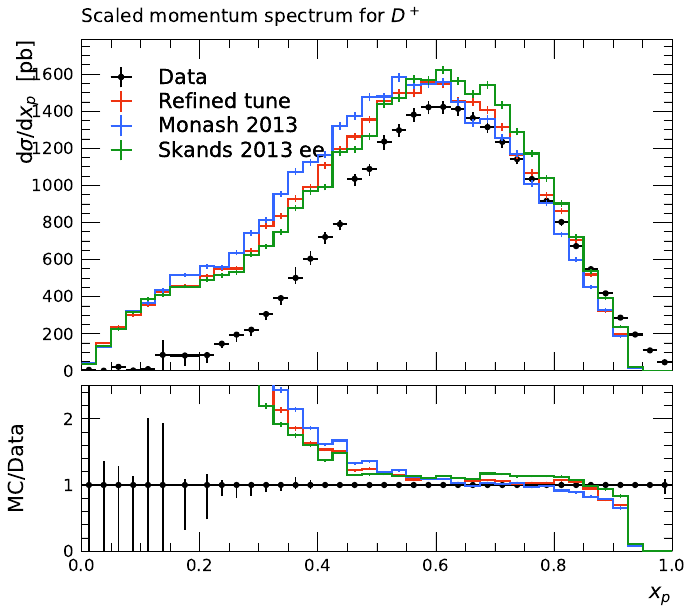}
\caption{$D^+$ scaled-momentum spectrum.}
\end{subfigure}\hfill
\begin{subfigure}[t]{0.48\textwidth}
\includegraphics[width=\linewidth]{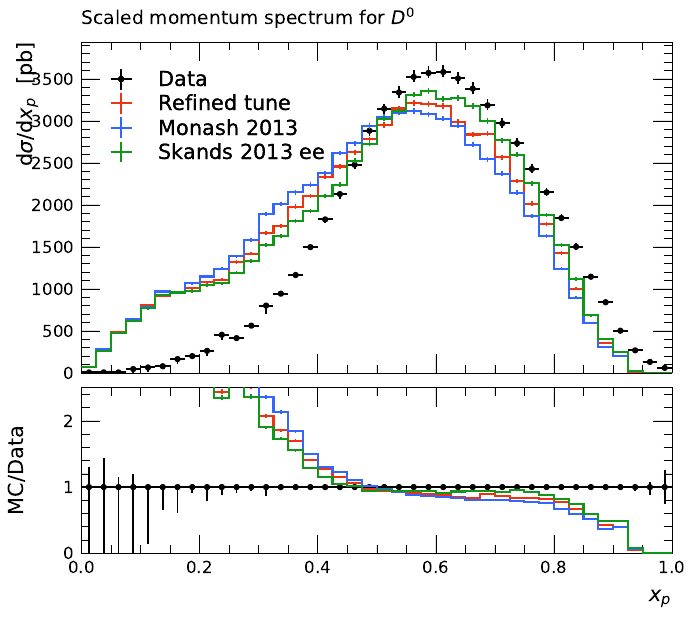}
\caption{$D^0$ scaled-momentum spectrum.}
\end{subfigure}
\vspace{0.8em}
\begin{subfigure}[t]{0.48\textwidth}
\includegraphics[width=\linewidth]{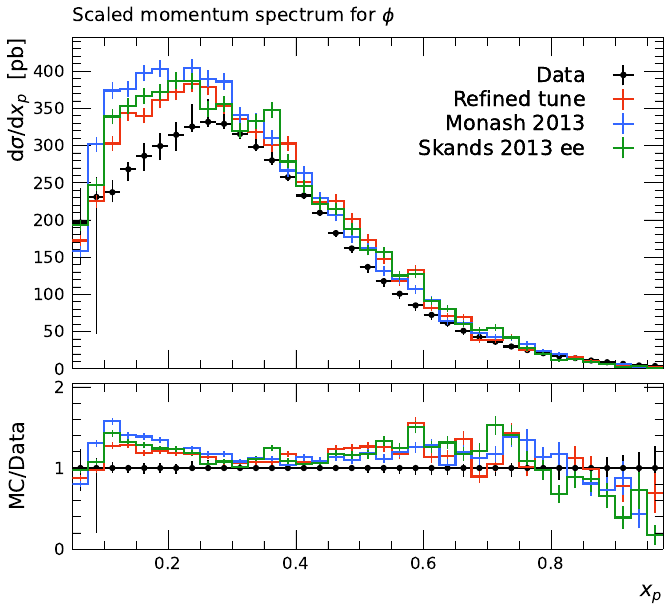}
\caption{$\phi$ scaled-momentum spectrum.}
\end{subfigure}\hfill
\begin{subfigure}[t]{0.48\textwidth}
\includegraphics[width=\linewidth]{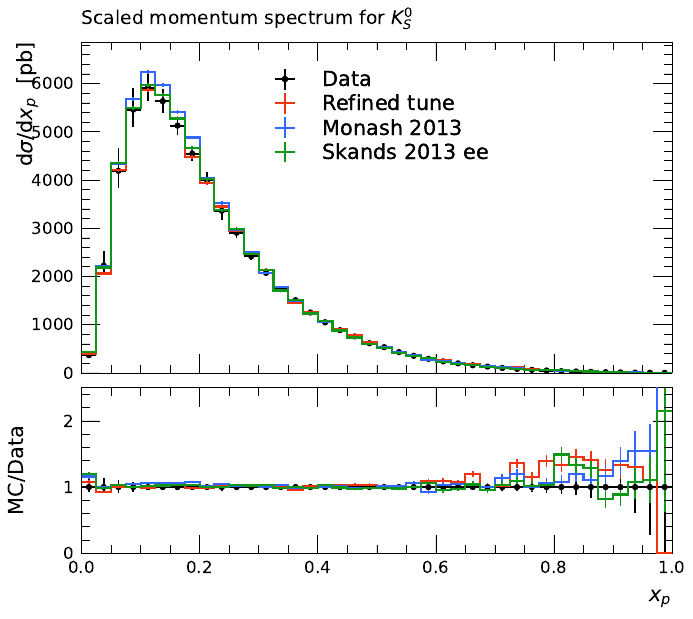}
\caption{$K^0_S$ scaled-momentum spectrum.}
\end{subfigure}
\caption{Selected BELLE~2025 meson spectra. The selected tune is closer to the data than Monash in some channels, although Skands gives the lowest aggregate BELLE~2025 score.}
\label{fig:belle2025-panels}
\end{figure}

The selected panels illustrate the data sets that control the comparison.
Figure~\ref{fig:low-energy-panels} shows the charged-hadron sector: BELLE~2013
favours the refined tune, whereas BABAR~2013 favours Skands. 
Figure~\ref{fig:belle2020-panels} shows representative BELLE~2020 single- and
dihadron spectra. This analysis carries most of the common-bin population and therefore dominates
the bin-weighted score. Figure~\ref{fig:belle2025-panels} shows representative
BELLE~2025 meson spectra, where Skands remains the best of the three
configurations. Additional panels are given in Appendix A, and
the full numerical breakdown is given in Appendix B.

\section{Discussion}
\label{sec:discussion}

The refined tune improves the combined BELLE/BABAR scores relative to Monash and Skands, but the improvement is not uniform across the five analyses. The improvement is small but reproducible within the scoring procedure used here. BABAR 2013 and BELLE 2025 still prefer the Skands reference, so the refined tune is not presented as a general replacement for existing \(e^+e^-\) tunes. The parameter changes also limit the interpretation. The basic Lund-shape
parameters remain close to the Monash values, while the added freedom is mainly
in flavour, vector-meson, diquark, isoscalar-meson, charm-vector, and
popcorn-baryon controls. The result is therefore consistent with changes mainly in the
species-composition and baryon-production balance of the PYTHIA string model,
rather than a large change in the longitudinal-fragmentation shape. The bin-weighted score is strongly weighted by BELLE~2020, which supplies 19,530 of the 20,803 bins in the common comparison. The bin-weighted ranking is therefore driven mainly by the single- and dihadron
block, not by a balanced average over the five analyses. For this reason, the
refined tune is best described as a channel-specific improvement on the selected
continuum scoring basis, with remaining differences in BABAR~2013 and BELLE~2025.

The residual pattern is also species dependent. In BABAR~2013, the refined tune is closest to the data for kaons, while Skands gives lower scores for pions and protons. In BELLE~2025, the largest tension is in the light-meson spectra, especially at low and high momentum; Skands remains lower than the refined tune for both light- and charm-meson groups. One-dimensional parameter scans show local responses to the string transverse-momentum width, vector-meson fractions, isoscalar suppression factors, and popcorn baryon parameter, but the available scan does not support a reliable covariance matrix or confidence interval for the tune parameters. The parameter values in Table 3 are reported to the precision supported by the profile study in Appendix C; the exact values used for event generation are retained in the origninal tune card.

\section{Conclusions}
\label{sec:conclusion}

We have presented a staged $e^+e^-$ refinement of a pp-derived PYTHIA~8 hadronization tune for $e^+e^-\to q\bar q$ production near $\sqrt{s}=10.52$--10.58~GeV. Starting from the five-parameter hadronization subset of the referenced pp tune, the refinement introduces additional flavour-sensitive controls in two steps guided by the remaining differences of the BELLE and BABAR data. The final comparison used 1,000,000 generated events per analysis and a fixed common set of 20,803 bins. The selected refined tune gives a bin-weighted score of 73.42, compared with 76.49 for the Skands $e^+e^-$ reference and 79.22 for Monash~2013 $e^+e^-$. It also gives the lowest equal-analysis and observable-family-balanced scores among the final compared tunes. It remains the best-scoring tune for BELLE~2013, BELLE~2017, and BELLE~2020, while Skands remains better for BABAR~2013 and BELLE~2025. These results define a small improvement for this data set rather than a general replacement for existing \(e^+e^-\) tunes. For the selected BELLE/BABAR data set, the extension of the pp-derived hadronization sector gives a small improvement in the bin-weighted score. This bin-weighted improvement is driven mainly by BELLE~2020 and does not remove the dataset-dependent differences, particularly in BABAR~2013 and BELLE~2025.

\section*{Acknowledgements}
This research work was supported by Princess Nourah bint Abdulrahman University Researchers Supporting Project number (PNURSP2026R106), Princess Nourah bint Abdulrahman University, Riyadh, Saudi Arabia.

\section*{Data availability}
The exact tune card and the numerical data underlying the figures and tables are available from the corresponding author upon reasonable request.

\section*{Conflict of interest}
The authors declare that they have no conflict of interest.

\appendix

\section{Selected supplementary comparison panels}
\label{app:panels}

The appendix collects supplementary comparison panels for the BELLE~2017 baryon sector and additional BELLE~2025 meson channels discussed in the main text.

\begin{figure}[p]
\centering
\begin{subfigure}[t]{0.48\textwidth}
\includegraphics[width=\linewidth]{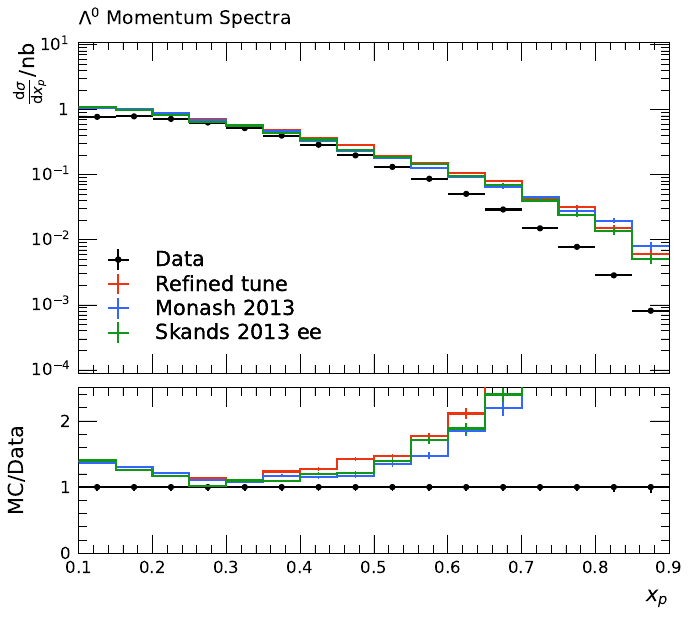}
\caption{$\Lambda^0$ momentum spectrum.}
\end{subfigure}\hfill
\begin{subfigure}[t]{0.48\textwidth}
\includegraphics[width=\linewidth]{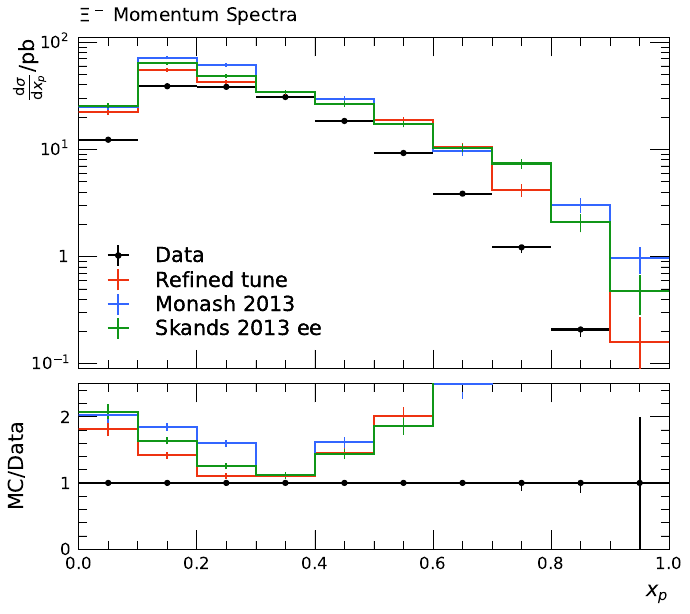}
\caption{$\Xi^-$ momentum spectrum.}
\end{subfigure}
\vspace{0.8em}
\begin{subfigure}[t]{0.48\textwidth}
\includegraphics[width=\linewidth]{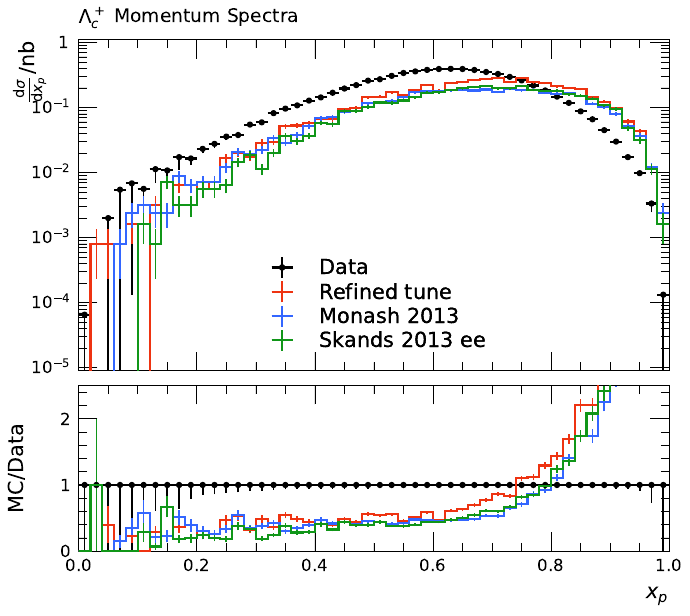}
\caption{$\Lambda_c^+$ momentum spectrum.}
\end{subfigure}\hfill
\begin{subfigure}[t]{0.48\textwidth}
\includegraphics[width=\linewidth]{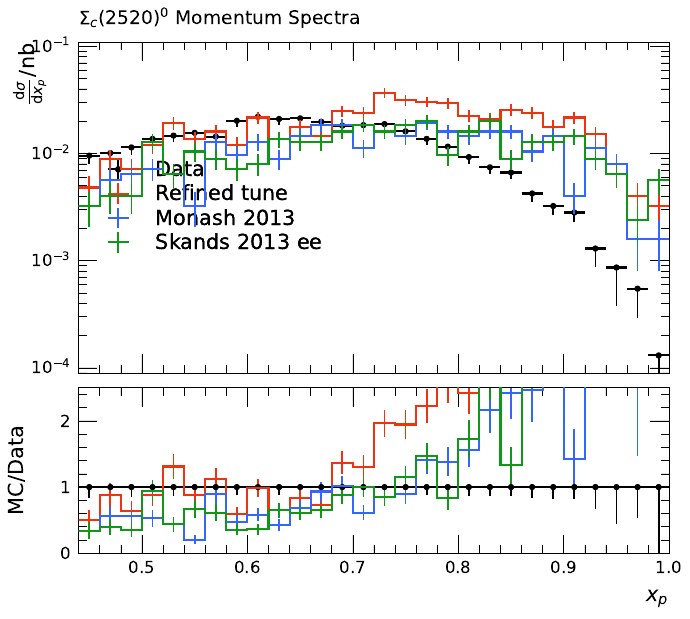}
\caption{$\Sigma_c(2520)^0$ momentum spectrum.}
\end{subfigure}
\caption{Selected BELLE~2017 baryon channels. The selected refined tune gives the lowest score for this baryon sample in the final comparison.}
\end{figure}

\begin{figure}[p]
\centering
\begin{subfigure}[t]{0.48\textwidth}
\includegraphics[width=\linewidth]{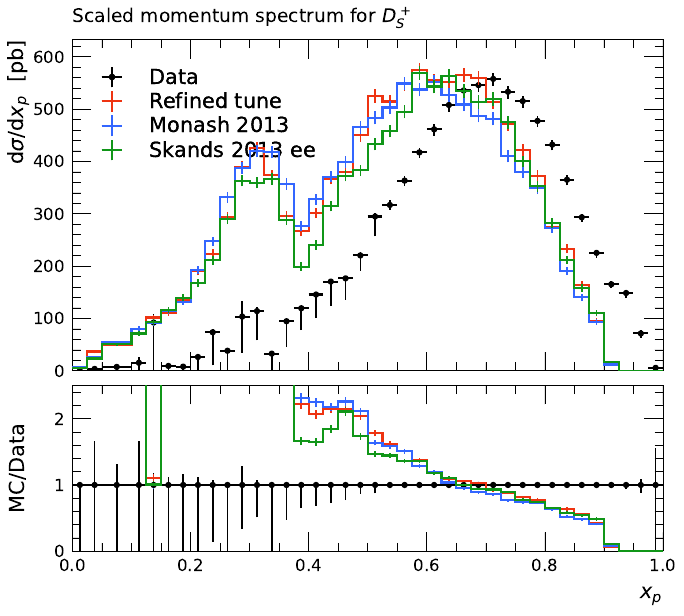}
\caption{$D_s^+$ scaled-momentum spectrum.}
\end{subfigure}\hfill
\begin{subfigure}[t]{0.48\textwidth}
\includegraphics[width=\linewidth]{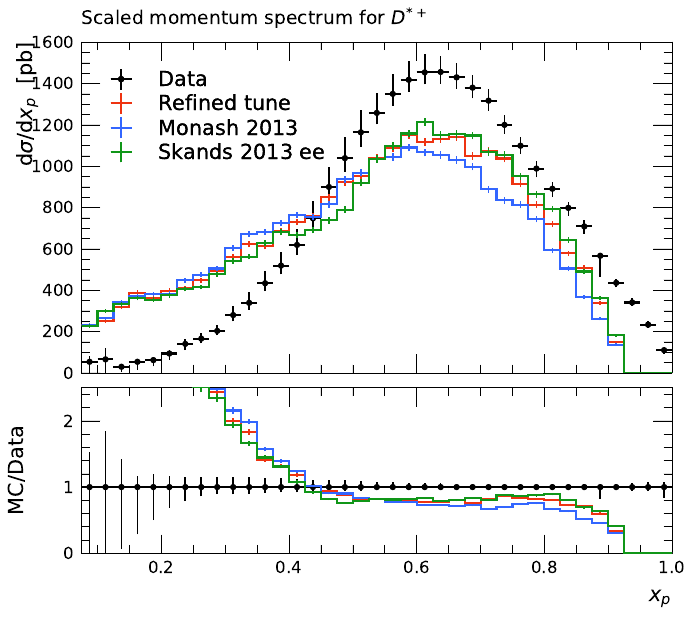}
\caption{$D^{*+}$ scaled-momentum spectrum.}
\end{subfigure}
\vspace{0.8em}
\begin{subfigure}[t]{0.48\textwidth}
\includegraphics[width=\linewidth]{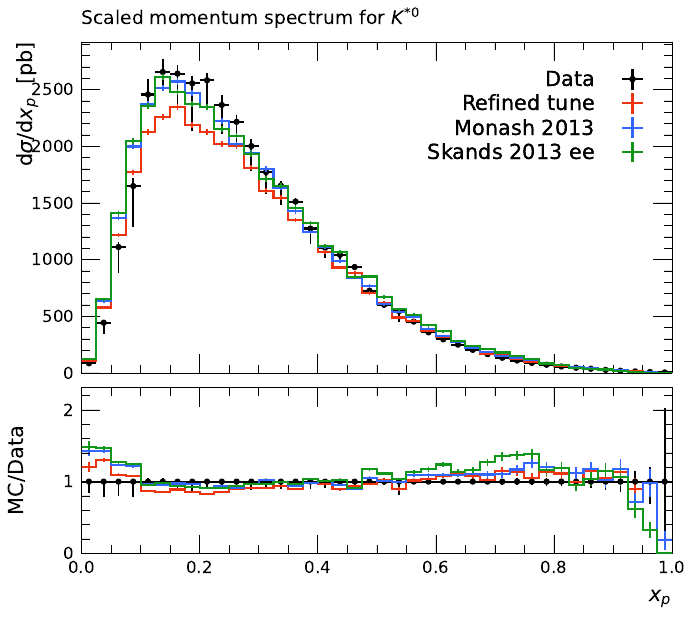}
\caption{$K^{*0}$ scaled-momentum spectrum.}
\end{subfigure}\hfill
\begin{subfigure}[t]{0.48\textwidth}
\includegraphics[width=\linewidth]{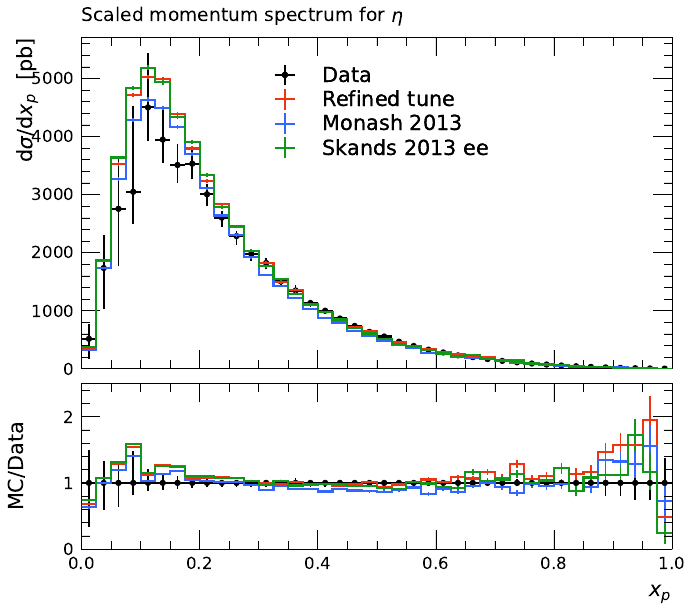}
\caption{$\eta$ scaled-momentum spectrum.}
\end{subfigure}
\caption{Selected supplementary BELLE~2025 channels. The reference Skands tune gives the strongest overall description of the BELLE~2025 meson spectra, while the selected refined tune remains competitive in several channels.}
\end{figure}

\clearpage
\section{Additional numerical table}
\label{app:numbers}
\setcounter{table}{0}
\renewcommand{\thetable}{\Alph{section}.\arabic{table}}

{\small
\setlength{\LTleft}{0pt}
\setlength{\LTright}{0pt}
\setlength{\tabcolsep}{4pt}
\renewcommand{\arraystretch}{1.08}

\begin{longtable}{@{}
>{\raggedright\arraybackslash}p{0.29\textwidth}
>{\raggedright\arraybackslash}p{0.18\textwidth}
r
r
r
@{}}
\caption{Full numerical breakdown of $\chi^2$, common-bin count, and bin-weighted score for the final three-way comparison.}
\label{tab:full-numeric}\\
\toprule
Analysis & Tune & $\chi^2$ & $N_{\mathrm{bin}}$ & Score \\
\midrule
\endfirsthead

\multicolumn{5}{@{}l}{\textit{Table \thetable\ continued.}}\\
\toprule
Analysis & Tune & $\chi^2$ & $N_{\mathrm{bin}}$ & Score \\
\midrule
\endhead

\midrule
\multicolumn{5}{r@{}}{\textit{Continued on next page}}\\
\endfoot

\bottomrule
\endlastfoot

\nolinkurl{BELLE_2013_I1216515} & Refined tune    &   38366.9 &   156 & 245.94 \\
                                & Monash~2013    &   45310.2 &   156 & 290.45 \\
                                & Skands $e^{+}e^{-}$ &   43205.8 &   156 & 276.96 \\
\addlinespace[2pt]

\nolinkurl{BELLE_2017_I1606201} & Refined tune    &   26092.8 &   266 &  98.09 \\
                                & Monash~2013    &   28240.4 &   266 &  106.17 \\
                                & Skands $e^{+}e^{-}$ &   28441.0 &   266 &  106.92 \\
\addlinespace[2pt]

\nolinkurl{BABAR_2013_I1238276} & Refined tune    &    8343.4 &   270 &  30.90 \\
                                & Monash~2013    &    6817.2 &   270 &  25.25 \\
                                & Skands $e^{+}e^{-}$ &    6578.5 &   270 &  24.36 \\
\addlinespace[2pt]

\nolinkurl{BELLE_2020_I1777678} & Refined tune    & 1431205.7 & 19530 &  73.28 \\
                                & Monash~2013    & 1545068.5 & 19530 &  79.11 \\
                                & Skands $e^{+}e^{-}$ & 1495701.7 & 19530 &  76.58 \\
\addlinespace[2pt]

\nolinkurl{BELLE_2025_I2849895} & Refined tune    &   23260.9 &   581 &  40.04 \\
                                & Monash~2013    &   22605.9 &   581 &  38.91 \\
                                & Skands $e^{+}e^{-}$ &   17396.2 &   581 &  29.94 \\
\addlinespace[2pt]

\textbf{Global total}           & Refined tune    & 1527269.7 & 20803 &  73.42 \\
                                & Monash~2013    & 1648042.3 & 20803 &  79.22 \\
                                & Skands $e^{+}e^{-}$ & 1591323.1 & 20803 &  76.49 \\

\end{longtable}
}


\section{Parameter sensitivity summary}
\label{app:sensitivity}
\setcounter{table}{0}

\begin{table}[p]
\centering
\caption{One-parameter sensitivity study around the selected refined tune. The tested range lists the sampled values for each parameter, while the remaining parameters were fixed at their selected values. The quantity \(\max |\Delta S_{\mathrm{analysis}}|\) is the largest absolute change in the equal-analysis score relative to the central tune within the tested range. All variations give ($\max|\Delta S_{\mathrm{analysis}}|>0.69)$, so the score changes are resolved within the profile scan. The listed ranges are the tested scan ranges only, not confidence intervals or correlated tune uncertainties.}
\label{tab:parameter-sensitivity}
\setlength{\tabcolsep}{3pt}
\renewcommand{\arraystretch}{1.08}

\begin{tabular}{@{}
>{\raggedright\arraybackslash}p{0.24\textwidth}
>{\raggedleft\arraybackslash}p{0.06\textwidth}
>{\raggedleft\arraybackslash}p{0.23\textwidth}
>{\raggedleft\arraybackslash}p{0.16\textwidth}
>{\centering\arraybackslash}p{0.21\textwidth}
@{}}
\toprule
Parameter &
Reported value &
Tested range &
\(\max |\Delta S_{\mathrm{analysis}}|\) &
Reported precision \\
\midrule

\nolinkurl{StringZ:aLund} &
0.75 &
0.6261--0.7816 &
6.00 &
2 significant digits \\

\nolinkurl{StringZ:bLund} &
1.0 &
1.0072--1.3864 &
5.89 &
2 significant digits \\

\nolinkurl{StringFlav:probStoUD} &
0.21 &
0.1964--0.2511 &
5.91 &
2 significant digits \\

\nolinkurl{StringPT:sigma} &
0.32 &
0.2796--0.3392 &
5.84 &
2 significant digits \\

\nolinkurl{StringFlav:probQQtoQ} &
0.092 &
0.0790--0.1037 &
6.62 &
2 significant digits \\

\nolinkurl{StringFlav:probSQtoQQ} &
0.77 &
0.7300--0.9875 &
6.24 &
2 significant digits \\

\nolinkurl{StringFlav:probQQ1toQQ0} &
0.042 &
0.0272--0.0523 &
6.03 &
2 significant digits \\

\nolinkurl{StringFlav:mesonUDvector} &
0.58 &
0.4507--0.6466 &
6.02 &
2 significant digits \\

\nolinkurl{StringFlav:mesonSvector} &
0.54 &
0.4805--0.6732 &
6.02 &
2 significant digits \\

\nolinkurl{StringFlav:mesonCvector} &
0.82 &
0.6847--1.1213 &
6.44 &
2 significant digits \\

\nolinkurl{StringFlav:etaSup} &
0.68 &
0.4310--0.8242 &
6.77 &
2 significant digits \\

\nolinkurl{StringFlav:etaPrimeSup} &
0.26 &
0.1161--0.2803 &
6.20 &
2 significant digits \\

\nolinkurl{StringFlav:popcornRate} &
0.32 &
0.2463--0.8896 &
6.49 &
2 significant digits \\

\bottomrule
\end{tabular}
\end{table}

\end{document}